\documentstyle[aps,preprint,epsf]{revtex}
\tightenlines
\begin{document}
\title{Two-Body Random Ensembles: From Nuclear Spectra to Random Polynomials}
\date{\today }
\author{Dimitri Kusnezov}
\address{ Center for Theoretical Physics,
Sloane Physics Laboratory,
Yale University, New Haven, CT 06520-8120}

\maketitle

\begin{abstract}
The two-body random ensemble (TBRE) for a many-body bosonic theory is
mapped to a problem of random polynomials on the unit
interval. In this way one can understand the predominance of  $0^+$
ground states, and analytic expressions can be
derived for distributions of lowest eigenvalues, energy gaps,
density of states and so forth. Recently studied nuclear
spectroscopic properties are addressed.
\end{abstract}

\hskip0.5cm


\vspace{0.5cm}

\narrowtext

The origins of spectroscopic properties of nuclei has received
renewed attention recently in the context of the two-body random
ensembles\cite{tbre-bos-b,tbre-fer-b,tbre-bos-c}. These studies provide an
understanding of which nuclear properties are robust, depending
only on the model space (1-- and 2-- body interactions), and which
depend on specific strengths  of interactions within the space.
The $0^+$ ground state is an example of a robust feature.
The starting point for analyses of TBREs are Hamiltonians of the
form
$H=\sum_{k}\varepsilon
_{k}c_{k}^{+}c_{k}+\sum_{ijkl}v_{ijkm}c_{i}^{+}c_{j}^{+}c_{k}c_{m}$,
where $c_k^+,c_k$ represent boson\cite{tbre-bos-a} or
fermion\cite{tbre-fer-a}  creation/annihilation
operators for a state $k$, and the coefficients $\varepsilon _{k},v_{ijkm}$
are taken as Gaussian random variables once certain physical constraints are
imposed, such that $H$ commutes with the generators of total spin, isospin
and so forth. The important distinction between the TBRE and the
conventional Gaussian orthogonal ensemble (GOE) 
description of many-body Hamiltonians is
that the latter does not include correlations between Hilbert
subspaces of different quantum numbers, which are essential to
understanding ground state and low energy spectroscopic
properties of nuclei. Although studied for some time, very
little is known analytically about the bosonic
TBRE\cite{tbre-bos-a,tbre-bos-b}.

We consider a bosonic model which can be treated
analytically, while retaining salient features of
more complex theories. The $U(4)$ Vibron  model\cite{il}
consists of  two type of 
bosons, $J^{\pi }=0^{+}, 1^{-}$, and
is used to describe the rotations and vibrations of diatomic
molecules. In contrast to the fermionic $U(4)$ problem of particles
in the $j=3/2$ shell, the Hilbert space of the bosonic theory
can be arbitrarily large. We will see that this model
describes many recently observed nuclear properties in the
$U(6)$ TBRE\cite{tbre-bos-b,tbre-bos-c}.
The Hamiltonian is\cite{il,fv}

\begin{eqnarray}
H &=&\frac 1N\left( \epsilon _ss^{+}s+\epsilon _pp^{+}\cdot \widetilde{p}
\right) +\left( \frac{c_0}2\left[ p^{+}p^{+}\right]
^{(0)}\cdot \left[ \widetilde{p}\widetilde{p}\right] ^{(0)}\right.  \nonumber\\
 && +\frac{c_2}2\left[
p^{+}p^{+}\right] ^{(2)}\cdot \left[ \widetilde{p}\widetilde{p}\right]
^{(2)} +\frac{u_0}2\left[ s^{+}s^{+}\right] ^{(0)}\cdot \left[ ss\right]
^{(0)}  \nonumber \\
&&+\frac{v_0}{2\sqrt{2}}\left( \left[ s^{+}s^{+}\right] ^{(0)}\cdot \left[ 
\widetilde{p}\widetilde{p}\right] ^{(0)}+h.c.\right)  \nonumber\\
 &&\left.  +\frac{u_1}2\left[
s^{+}p^{+}\right] ^{(1)}\cdot \left[ \widetilde{s}\widetilde{p}\right]
^{(1)}\ \right)/N(N-1) \label{hvib} 
\end{eqnarray}
where $s^{+}(s)$ and $p_\mu ^{+}(\widetilde{p}_\mu =-p_{-\mu })$ are the
spherical-tensor creation (annihilation) operators for states with $J^\pi
=0^{+}$ and $1^{-}$( projection $\mu =0,\pm 1),$ respectively. The square
brackets indicate angular momentum couplings and dots scalar products.
Since 
the matrix elements of the 1- and 2-body interactions are proportional to $N$
and $N(N-1)$,  scaling allows all coefficients to be Gaussian random
numbers of unit variance. The random Gaussian variables are grouped into a
vector $x=(c_0,c_2,u_0,u_1,v_0,\epsilon _s,\epsilon _p)$. The matrix
elements of $H$ are well known in the vibrational   basis $\left|
Nn_pJm\right\rangle $,
where $N$ is the total number of bosons, $n_p=0,1,...,N$ is the number of 
$J^\pi =1^{-}$ bosons, $J=n_p,n_p-2,...,1$ or 0 is the total angular
momentum of the many-body state ( $m=-J,...,J$
is omitted
since it adds a trivial degeneracy.) We also note that
even (odd) $J$ states have only even (odd) values of $n_p$, resulting in an
odd/even effect depending on the choice of $N$. While this effect is easily
treated, we will focus only on even $N$ to simplify the presentation.

When the dimension of the Hilbert
space is large, which is typical of many-body problems, a common
approach to diagonalization is the Lanczos method. Here
successive iterations of the Hamiltonian on an arbitrary trial
wavefunction $\Psi _0$ are performed to reduce the Hamiltonian to a
tridiagonal form. For a given spin $J$ of an $N$
-body state, we label the basis by $n_p=J,J+2,...,N-1$ or $N$. Clearly the
dimension of the $J=0$ Hilbert space is maximum, 
while that of $J=N$ is minimum. While it
has been {\em argued} for some TBRE's that the dimensionality or
the width of the lowest eigenvalue distribution is responsible
for the preponderance of $0^+$ ground states, we will see that this is not
the origin here. 

We choose the trial Lanczos state to be that with
$n_p=J$. Enumerating the states by an index $k=(n_p-J)/2$, the
Hamiltonian (1) assumes the tridiagonal form $H\Psi _k=\beta _k\Psi _k+\alpha
_{k-1}\Psi _{k-1}+\alpha _{k+1}\Psi _{k+1}$ where $\alpha _k,\beta _k$ are
the matrix elements of $H$ in this basis. The study of tridiagonal matrices
is intimately linked to orthogonal polynomials and their recursion
relations, where $k$ might represent the order of the
polynomial\cite{ismail}. In particular, the family of
recursively generated, real polynomials are
related to  the characteristic polynomial $D_i(E)=\det ({\cal H}
_i-E\cdot 1_i)$ of the $i\times i$ tridiagonal matrix  ${\cal H}_i$.
Consequently the zeros of the polynomials $D_i$ are related to the
eigenvalues of  ${\cal H}_i.$ There are several theorems which
have developed bounds for the zeroes of $D_i$ as well
as expressions for the extreme eigenvalues which we
can use to derive properties of the TBRE\cite{ismail,vandoorn}. 

We consider two cases here: large $N$ and $N=2$.
For $N=2$, the order of the
interaction (2-body) is equal to the number of particles, and we expect to
recover the GOE. The only
allowed states are $J=0,1,2$. The $J=1,2$ Hilbert spaces are 1-dimensional
while $J=0$ is 2-d. It can be readily checked that the density of $J=0$
states gives the Wigner semi-circle (i.e. 
the $2\times 2$ matrix can be expressed as one with
GOE measure), while the $J=1,2$ are always Gaussian.
The same is true for $J=J_{\max }=N$ for any $N$, since the Hilbert space
for the maximum spin states is always one-dimensional so that the density of
states is purely Gaussian (Fig. 1(b)).

 In the following, we will use the large $N$ limit
(typically in molecules $N\sim 100$\cite{il,fv}; but $N=8$ calculations already
agree with our analytic predictions below). 
We next define $z=n_p/N$ and $j=J/N$
and construct the functions $\alpha (z,j)$, $\beta (z,j)$ from the analytic
matrix elements of the tridiagonal matrix,
$\beta (z,j) = \gamma_1 z^2+\gamma_2 z+\gamma_{3,j}$, 
$\alpha (z,j) =\gamma_4 z(1-z)+o(1/N)$
where $\gamma_k$ are linear combinations of the random coefficients
$x$. $\alpha$ and $\beta$ are the off-diagonal and diagonal
matrix elements, respectively. Then, in the large 
$N$ limit, the lowest eigenvalue for each spin has the form
\cite{ismail,hollen} 
\begin{equation}
E_{\min }^j=\inf_z(f_j(z)),\quad f_j(z)=\beta (z,j)-2|\alpha (z,j)|.
\end{equation}
In terms of the parameters of $H$ in Eq. (1) and the matrix
elements of the interactions\cite{il,fv}, we can express
$f_j(z)=az^2+bz+d_j$, where $z\in [j,1]$ and
\begin{eqnarray}
a&=&r_1\cdot x=(\frac{c_0}{6}+\frac{c_2}{3}+\frac{u_0}{2}-u_1
  +\frac{|v_0|}{\sqrt{6}})\frac{N}{N-1},\nonumber\\
b&=&r_2\cdot
x=\varepsilon_p-\varepsilon_s+(u_1-u_0-|v_0|/\sqrt{6})
\frac{N}{N-1},\\
d_j&=&r_3\cdot x=\varepsilon _s+u_0/2+j^2(c_2-c_0)/6.\nonumber
\end{eqnarray}
To check this approach, consider only the off-diagonal
interactions ($\beta =0$). Then
we expect for small $J$ that the minimum energy approaches
$E_{\min }^j=-|v_0| N/4\sqrt{6}(N-1)$. The numerical value,
denoted $E_{calc}^j$, can be seen to converge to this result. 
For $J=0,1,2$,  we $E_{calc}^j/E_{\min}^j=1.06,1.05,1.03$ 
for $N=20$; $1.01,1.01,1.01$ for
$N=80$. Hence the desired spectral properties of
$H$ can be recast in terms of random polynomials on the unit
interval.
This should be generally true for any
bosonic or fermionic theory since the Lanczos approach can be applied to
either. The main effort is to determine whether
there is any approximate analytic behavior of the functions
$\alpha ,\beta $, although it is suggested to be generally true\cite{hollen}.

Many general properties of random or Kac polynomials are known, but
typically for higher order functions which are otherwise unrestricted\cite
{kac}. To understand the properties of this TBRE, we first compute the
distribution of coefficients.  Since the variables $x$ are taken as Gaussian
with measure $P(x)\propto \exp [-\sum_kx_k^2/2]$, one can compute the
distribution of coefficients of the random polynomial with
${\cal P}(a,b,d_j)\propto
\int dxP(x)\delta (a-r_1\cdot x)\delta (b-r_2\cdot x)\delta
(d_j-r_3\cdot x)$.  Integrating yields: 
\begin{equation}
{\cal P}(a,b,d_j)\propto 
\exp [-\frac 12{\cal J}^TM^{-1}{\cal J}],\qquad {\cal J} =(a,b,d_j).
\end{equation}
Here $M_{\alpha \beta }^{-1} = (2\det M)^{-1} \epsilon _{\alpha ab}\epsilon
_{\beta cd} (\delta _{ik}\delta _{jl}-\delta _{ij}\delta
_{kl}) r_{a,i} r_{b,j}$ $r_{c,k} r_{d,l}$ is the inverse of 
$M_{\alpha \beta }=r_\alpha \cdot r_\beta$,
and $r_{a,i}$ is the $i^{th}$ component of $r_a$.

As in the $U(6)$ model\cite{tbre-bos-b}, the frequency of a ground state 
$0^+$ is approximately 70\% (for even $N$; 70.5\%,
71.7\%, 72.3\%, 72.5\% for $N=8,16,32,64$). To
understand this, we first compute where the minima $z=z_0$ of
$f_0(z)$ are located. Integrating the location
of the minima over the distribution of
coefficients (4), we find the distribution of minima to be:

\begin{equation}
P(z_{0})=\frac{1}{2\pi }
\frac{\sqrt{r_{1}^{2}r_{2}^{2}-(r_{1}\cdot r_{2})^{2}}}{
  r_{1}^{2}z_{0}^{2}+(r_{1}\cdot r_{2})z_{0}+ r_{2}^{2}/4}.
\end{equation}
In Fig. 1(a) we compare this function to results from $10^6$
numerical diagonalization of Eq. (1) for 
N=8,16,32 and 64 bosons. We estimate the frequency
of $0^+$ ground states by first understanding where in $z\in
[0,1]$ are the minima located. For 50\% the cases, $a<0$, so
$f_0(z)$ is an inverted parabola, and its minimum is at $z=0$ or
1, each occurring with equal probability (25\%). For the
remaining 50\% of the
cases, $a>0$. For these cases, the probability of having the minimum in
$z\in (0,1)$ is
\begin{equation}
\frac{1}{2}\int_{0}^{1}P(z_0)dz_0=\frac{1}{2\pi }
{\rm tan}^{-1}\left[ \frac{2\sqrt{r_{1}^{2}r_{2}^{2}-(r_{1}\cdot r_{2})^{2}}
}{r_{2}^{2}+2r_{1}\cdot r_{2}}\right],
\end{equation}
which has a value of 22\%,
leaving 14\% at $z=0$ and 14\% at $z=1$. 
Next we ask when the $0^+$ state is a
global minimum (over all $J$). At $z=0$, only $0^+$ states are
allowed (since a state $J>0$ has $n_p\geq J$, hence $z>0$).
Hence for at least 39\% of the cases, $0^+$ is the ground
state. At $z=1$, all even $J$ are allowed, and from (2)-(3), 
$E_{min}^j-E_{min}^0=j^2 (c_2-c_0)/6$. Since $c_0,c_2$ are
Gaussian random numbers, half the time $0^+$ is the ground
state ($E_{min}^0<E_{min}^j$), and the other 
half, the state of maximum $j$, or $J=N$.
So at $z=1$, 19.5\% of the ground states are $0^+$ and 19.5\% are
$J=N$. For $z\in (0,1)$, we note that $J=0$ has even $n_p$ and
hence is only allowed at half the points. Consequently of the
22\% of the minima here, no more than 11\% can be $0^+$ ground
states. Hence we estimate a roughly 69.5\% frequency of $0^+$
ground states, in very good agreement with observations.
 
Since it is more likely to find the minima of $f_j(z)$ on the 
boundary than inside, we can compute distribution functions of
interest by restricting attention to the edges $z=j$ and
$z=1$. While this is approximate, it does yield predictions which
agree very well with the molecular $U(4)$ and nuclear $U(6)$ TBREs. Averaging
inf${}_z(f_j(z))$ over $P(a,b,d_j)$ yields the 
distribution of lowest eigenvalues, denoted $p_J(E)$. These have the form
\begin{eqnarray}
p_{0}(E) &\propto  &\exp (-E^{2}/2r_{3}^{2})
{\rm erfc}(EA_{1}) \nonumber \\
  & & +A_{3}\exp (-E^{2}/2R^{2})
{\rm erfc}(EA_{2})  \label{emin} \\
p_{J_{\max }}(E) &\propto  &\exp (-E^{2}/2R^{2})
\end{eqnarray}
where $R^2=(r_{1}+r_{2}+r_{3})^{2}=1.07$, $A_{1}=(r_{1}\cdot r_{3}+r_{2}\cdot
r_{3})/(hr_{3})=0.663,$ $A_{2}=(R^{2}-r_{1}\cdot r_{3}-r_{2}\cdot
r_{3})/hR=0.632,$ $A_{3}=hr_{3}/(R\sqrt{2\det M})=0.423,$ and $%
h^{2}=2(r_{3}^{2}(r_{1}^{2}+2r_{1}\cdot r_{2}+r_{2}^{2})-(r_{1}\cdot
r_{3}+r_{2}\cdot r_{3})^{2}),$ (evaluated at $N=64$). 
For other values of $J$ one can readily derive the general
form which is a linear combination of two terms similar to (\ref{emin}).
In Fig. 2(a) we compare (\ref{emin}) (solid) to a Gaussian
(dashes) and to results from diagonalization of $H$ for selected
$N$. The low energy excess is readily described, and the results
are not Gaussian for any $N$, in contrast to results from the
dilute limit\cite{tbre-bos-a}. In 2(b) we
compare the same function with slightly modified parameters 
($r_3=\sqrt{2}$ and $A_1=3/4$) and see that this functional form
can readily account for the observed asymmetry in the $U(6)$
results\cite{tbre-bos-b}. The results for maximum $J$ are shown
in Fig. 1(b). Since this Hilbert space is 1-d, the distribution
also corresponds to the level density. 

It is interesting to use $f_j(z)$ to estimate
the level density $\rho_J(E)$
for states of spin $J$ by averaging this over the space of
random polynomials. For $J=J_{max}$, we trivially recover (8).
For $J=0$ we find
\begin{equation}
\rho_0(E) =  \int_{0}^{1}dz\frac{1}{\sqrt{2\pi g(z)}}\exp \left[ -\frac{
E^{2}}{2g(z)}\right]
\end{equation}
where $g(z) ={\cal Z}^{T}M{\cal Z}$, ${\cal Z}=(z^{2},z,1)$,
which agrees well with calculations (Fig. 3). 
$\rho_0(E)$ has moments: 
\begin{equation}
\langle E^{2n}\rangle_{J=0} =\frac{2^{n}}{\pi }\Gamma
(n+1/2)\int_{0}^{1} g(z)^{n} dz.
\end{equation}
One can see that the shape of the level density is
a superposition of
Gaussians of varying width.
The width of the $0^+$ density of states is $\langle
E^{2}\rangle_{J=0} =0.784$, while for the maximum 
spin $\langle E^{2}\rangle_{J=N} =1.137$. 
In this model we see that there is no direct relation between the
widths of the distributions and the probability of having a
$0^+$ ground state, as has been conjectured in other
TBREs\cite{tbre-fer-b,tbre-bos-b}. Certainly the functional
dependence of these moments are distinct from the dependence of
(5)-(6), which is related to the ground state problem.
Hence, the level densities do not reflect
ground state properties in the sense that correlations between
subspaces of different $J$, which are central to that question, 
are not reflected in these functions.

The distribution of $1^--0^+$ energy gaps, 
$\tilde{g}=E_{min}^{1/N}-E_{min}^0$, denoted ${\tt p}(g)$, 
is obtained by averaging $\tilde{g}$ over Eq. (4):
\begin{eqnarray}
{\tt p}(g)&\propto&  \exp (-g^{2}/4r_{2}^{2})\left[ B_{1}\exp
  (-g^{2}(r_1^2+r_1\cdot r_2)/\Delta)
\right.\\
& &\left. \times {\rm erfc}(EB_{2}) - {\rm erfc}(gB_{3})\right]\nonumber
\end{eqnarray}
where we find the scaling $g=\sqrt{2} N \tilde{g}$. Here,
$B_{1}=r_{2}/\sqrt{r_{2}^{2}+4r_{1}\cdot r_{2}+4r_{1}^{2}}  
=1.059$,$B_{2}=(2r_1^2+r_1\cdot
r_2-\Delta)/2\sqrt{d\Delta}=-0.03$,
$B_{3}=(r_{1}\cdot
r_{2}+r_{2}^{2})/(2r_{2}\sqrt{d})=-0.311$,
$d=r_1^2r_2^2-(r_1\cdot r_2)^2$ and $\Delta = 4(r_1^2+r_1\cdot
r_2)+r_2^2$ (the numerical value is for $N=64$). 
This scaling in $N$ is evident
in Fig. 4(a) when we plot versus $g$. 
We note that ${\tt p}(g)$ describes the shape away
from the origin. (Near the origin there is an abundance of small
gaps that arise from the omitted region $z\in (0,1)$).
 We use the same function in
Fig. 4(b) to compare to the $E_{2_1^+}-E_{0_1^+}$ gaps computed
in the IBM, with $g= N \tilde{g}$. The same scaling is apparent.
Finally, the distribution of
$R_{4/2}=(E_{4^+}-E_{0^+})/(E_{2^+}-E_{0^+})$
has been measured in the $U(6)$ TBRE. The analogous quantity in this model is
$R_{2/1}=(E_{2^+}-E_{0^+})/(E_{1^-}-E_{0^+})$. The 
distribution $P(R_{2/1})$ can  be derived and has peaks at 2
and 3 ($=(J+1)/(J-1)$ for $J=2$), consistent with those in the
$U(6)$ model. To leading order this function is just a sum of
two delta functions, and higher order corrections must be
included to get the shape.

We have examined a bosonic TBRE which shares the salient features of more
complex TBREs, such as a Wigner limit for small $N$, Gaussian
level densities for certain states, preponderance of $0^+$
ground states, and so forth. By mapping the TBRE onto random
polynomials on the unit interval, we are able to analytically
understand many properties of the TBRE found
numerically, including the frequency of $0^+$ ground
states. We find the latter is not attributed to the width of the
level densities or the dimension of the Hilbert spaces. Rather,
the various interactions in $H$ tend to put the extreme values
of spin ($J=0,N$) at
the ends of the spectra, enhancing their chances to be the
ground state. These results 
provide the first analytic understanding of
ground state properties, distributions of lowest eigenvalues,
gaps and level densities for the TBRE.
These functions are also found to describe the
nuclear properties obtained in the IBM. Scaling behavior has
also been predicted and verified in certain observables. Since
the analytic Lanczos approach can be generally applied,
it would be interesting to see if a more general connection can
be made between the TBREs and random polynomials\cite{kbf}. 

I thank A. Frank, R. Bijker, R. Casten
and F. Iachello for many interesting discussions.
Work supported by the U.S. DOE contract DE-FG02-91ER-40608.

\begin{figure} 
\begin{center}
    \leavevmode
  \epsfysize=7.4cm\epsfbox{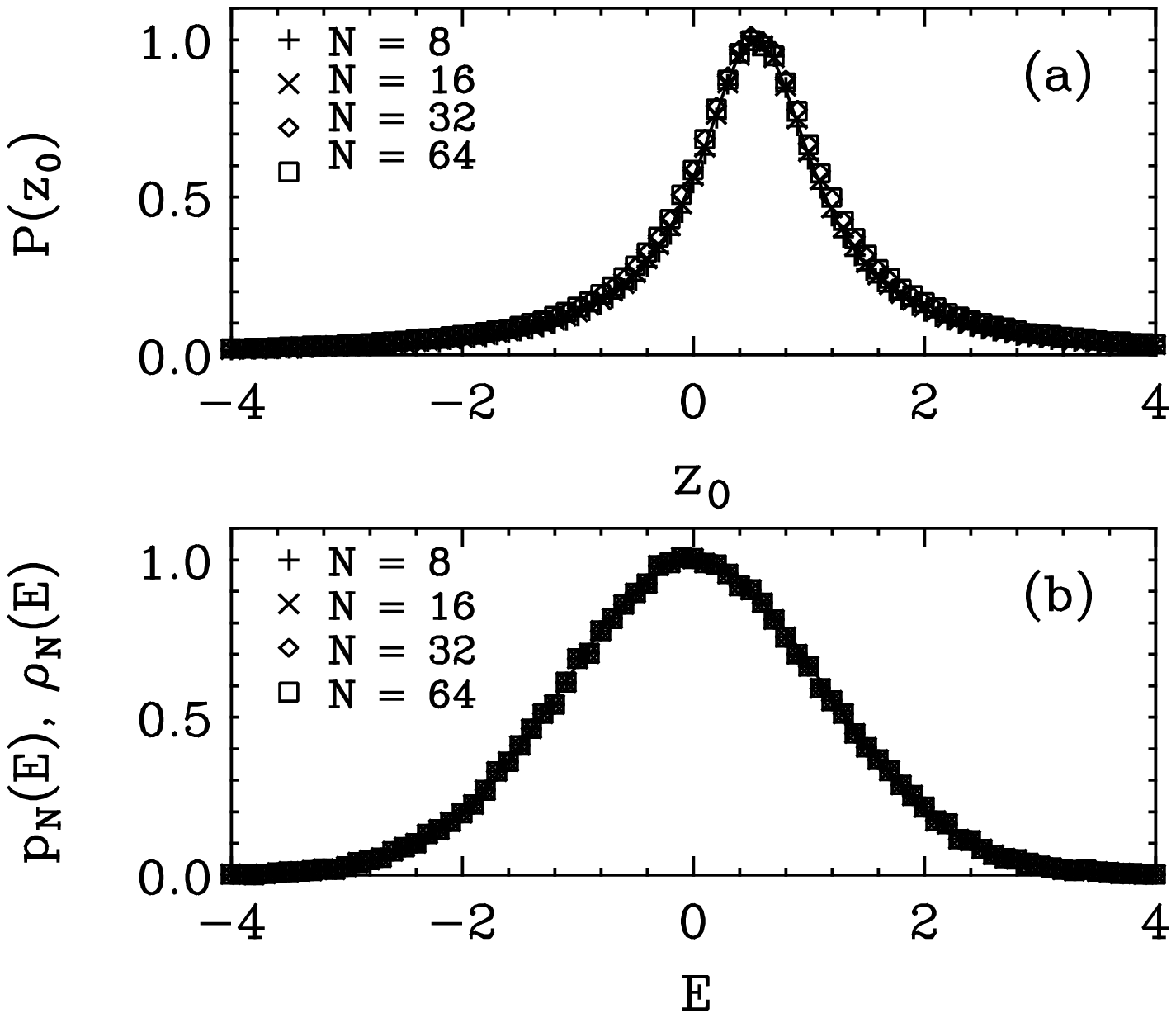}
  \caption{(a) Theoretical distribution (5) of extrema of $f_0(z)$
  (solid) compared to 
  numerical TBRE results (symbols). (b) Theoretical eigenvalue
  distribution for highest spin
  states $p_N(E)$ (solid; Eq. (8)) are seen to be Gaussian and agree with
  numerical results  (symbols). This is also the level density
  $\rho_N(E)$ of $J=N$ states
  for any $N$. (In this and the following figures, the 
  vertical scale is arbitrary).}
  \label{fig:one}
  \end{center}
\end{figure}

\begin{figure}
  \begin{center}
    \epsfxsize=8.cm\epsfbox{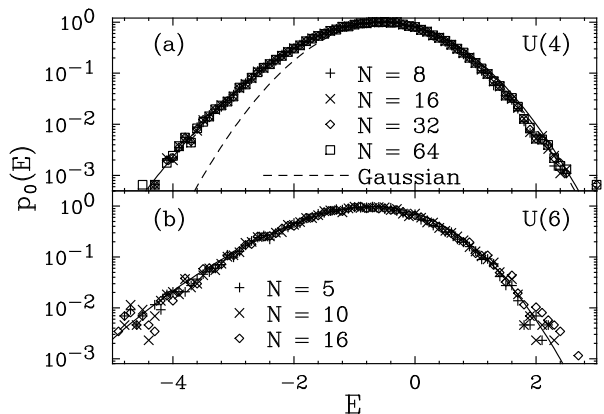}
    \caption{(a) Lowest eigenvalue distribution (7) for $J^\pi=0^+$ states
      (solid) compared to numerical TBRE calculations. The
      Gaussian is for reference. (b) Same as (a) but for the
      nuclear case.}
    \label{fig:thr}
  \end{center}
\end{figure}

\begin{figure}
  \begin{center}
    \epsfxsize=8.cm\epsfbox{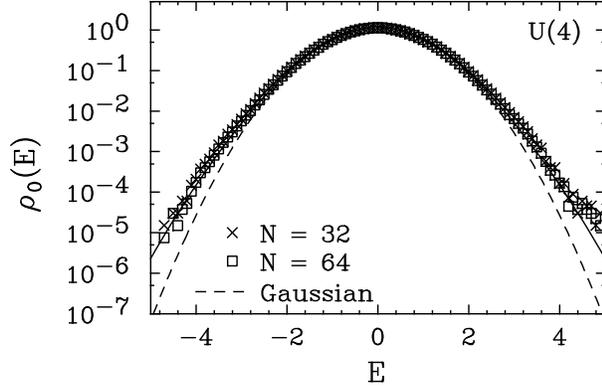}
    \caption{Theoretical density of states (9) for $0^+$ states
      (solid) compared to a
      Gaussian, and numerical TBRE calculations for selected $N$
      (symbols).}  
    \label{fig:two}
  \end{center}
\end{figure}
 
\begin{figure}   

  \begin{center}
    \epsfxsize=8.cm\epsfbox{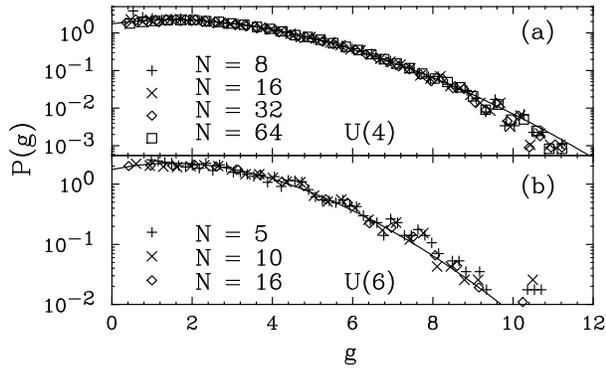}
    \caption{(a) Distribution of energy gaps ${\tt p}(g)$ where $g=\sqrt{2}N
      (E_{min}(J=1)-E_{min}(J=0))$. The description is good,
      demonstrating a predicted scaling, with the exception of
      the behavior at the origin. (b) Same as (a) but for the
      nuclear case. Here  $g=N (E_{min}(J=2)-E_{min}(J=0))$.}
    \label{fig:fou}
  \end{center}
\end{figure}
 
\end{document}